\documentclass[aps,prb,twocolumn,nowpacs,superscriptaddress,groupedaddress,amsmath,amssymb]{revtex4}
\usepackage {graphicx,epsfig,graphics,color}

\begin{document}


\title{Measurement of the magnetic penetration depth of a superconducting MgB$_{2}$
thin film with a large intraband diffusivity}

\author{Jeehoon Kim}
\email[Corresponding author: ]{jeehoon@lanl.gov} \affiliation{Los
Alamos National Laboratory, Los Alamos, NM 87545}
\author{N. Haberkorn}
\affiliation{Los Alamos National Laboratory, Los Alamos, NM 87545}
\author{Shi-Zeng Lin}
\affiliation{Los Alamos National Laboratory, Los Alamos, NM 87545}
\author{L. Civale}
\affiliation{Los Alamos National Laboratory, Los Alamos, NM 87545}
\author{E. Nazaretski}
\affiliation{Brookhaven National Laboratory, Upton, NY 11973}
\author{B. H. Moeckly}
\affiliation{Superconductor Technologies Inc., Santa Barbara, CA
93111}
\author{C. S. Yung}
\affiliation{Superconductor Technologies Inc., Santa Barbara, CA
93111}
\author{J. D. Thompson} \affiliation{Los Alamos National
Laboratory, Los Alamos, NM 87545}
\author{R. Movshovich}
\affiliation{Los Alamos National Laboratory, Los Alamos, NM 87545}

\date{\today}

\begin{abstract}

We report the temperature dependent magnetic penetration depth
$\lambda(T)$ and the superconducting critical field $H_{c2}(T)$
in a 500-nm MgB$_{2}$ film. Our analysis of the experimental
results takes into account the two gap nature of the
superconducting state and indicates larger intraband diffusivity
in the three-dimensional (3D) $\pi$ band compared to that in the
two-dimensional (2D) $\sigma$ band. Direct comparison of our
results with those reported previously for single crystals
indicates that larger intraband scattering in the 3D $\pi$ band
leads to an increase of $\lambda$. We calculated $\lambda$ and the
thermodynamic critical field $H_{c}\approx$2000 Oe employing the
gap equations for two-band superconductors. Good agreement
between the measured and calculated $\lambda$ value indicates the
two independent measurements, such as magnetic force microscopy
and transport, provide a venue for investigating superconducting
properties in multi-band superconductors.

\end{abstract}

\maketitle

\section {Introduction}

During the last decade a significant effort has been made to
understand the mechanism of two-band superconductivity in
MgB$_{2}$.\cite{Nagamatsu,Gurevich2003,Kogan2009,Welp} MgB$_{2}$
has two s-wave gaps residing on four different disconnected Fermi
surface (FS) sheets: two axial quasi two-dimensional (2D)
$\sigma$-band sheets and two contorted three-dimensional (3D)
$\pi$-band sheets. The $\sigma$ band forms two concentric
cylindrical sheets via in-plane sp$^{2}$ hybridization of the
boron valence electrons. The $\pi$ band results from the strongly
coupled covalent bonding and antibonding of the boron $P_{z}$
orbitals.\cite{Samuely} Multiple bands allow for both inter- and
intra-band scattering. It is thus possible to tune the upper
critical field ($H_{c2}$) via doping, which has different effects
on the inter- and intra-band scattering
strengths.\cite{Gurevich2003,Senkowicz,Matsumoto,Braccini} In
MgB$_{2}$ the anisotropy of the temperature dependent penetration
depth $\lambda$,
$\gamma_{\lambda}(T)=\lambda_{c}(T)/\lambda_{ab}(T)$ shows
remarkably different behavior compared to that of $H_{c2}$,
$\gamma_{H_{c2}}(T)=H^{ab}_{c2}(T)/H^{c}_{c2}(T)$.\cite{Kogan2002,Fletcher}
This difference indicates that the two-band nature of
superconductivity profoundly alters the superconducting
properties compared to those in a single band
material.\cite{Tinkham} For example, the equations for critical
fields and depairing current  as a function of $\lambda$ and
$\xi$ should be modified due to the inter/intra-band scattering.
Knowledge of the absolute values of $\lambda$ and $\xi$ is also
important for technological applications.\cite{Collings} For
example, the acceleration field in superconducting radio
frequency (SRF) cavities could be enhanced by covering
conventional superconducting Nb cavities with
superconductor/insulator multilayers (such as MgB$_{2}$) with
higher thermodynamic critical field ($H_{c}$).\cite{Gurevich2006}

A number of measurements have been performed to determine the
absolute value of $\lambda$ in
MgB$_{2}$.\cite{Kogan2009,Kogan2002, Fletcher} The reported
values of $\lambda$ range from 40 nm to 200 nm, indicating that
$\lambda$ is strongly affected by inter- and intra-band
scattering.\cite{xxx1, Golubov1, Dahm, Chen, Lee, Simon,
Finnemore} In this paper we present measurements of the absolute
values of $\lambda(T)$, employing low temperature magnetic force
microscopy (MFM), and of the angular-dependent $H_{c2}(T,\theta)$
performed via electrical transport, in a 500-nm thick MgB$_{2}$
film. Our MgB$_{2}$ film can be described by the dirty limit
two-band Usadel equations. We analyze the measured values of
$H_{c2}$ and $\lambda$ using a model developed for dirty
superconductors,\cite{Gurevich2003} which simplifies the analysis
compared to that reported in Ref. 15. We investigate
theoretically the influence of the inter/intra-band scattering on
the superconducting properties. Using a two-band superconductor
model with parameters obtained from a fit to $H_{c2}(T,\theta)$,
we calculate $\lambda$ and $H_{c}$ which are consistent with the
experimental values.


\section {Experiment}

A MgB$_{2}$ film was grown on a $r$-sapphire substrate by a
reactive evaporation technique.\cite{Moeckly 2006,Gu 2006} The
film is epitaxial and shows columnar growth morphology, with the
$c$ axis tilted by a few degrees from the normal direction of the
substrate. For more details see Ref. 22. The sample has
dimensions L=4 mm $\times$ W=5 mm $\times$ t=500 nm, and exhibits
a full superconducting volume fraction based on measurements
using a commercial SQUID magnetometer (Quantum Design magnetic
property measurement system, MPMS) All MFM measurements described
here were performed in a home-built low temperature MFM
apparatus.\cite{NazaretskiRSI2009} Temperature dependent vortex
images were taken in the frequency-modulated mode after a small
magnetic field was applied above $T_{c}$ (field-cooled). We used
high resolution SSS-QMFMR cantilevers.\cite{Nanosensors} The
magnetic field was always applied perpendicular to the film
surface and parallel to the MFM tip. The absolute values of
$\lambda(T)$ were determined by comparing the Meissner response
curves with those for a reference sample at 4
K.\cite{lambda,Jeehoon} The Meissner technique for the $\lambda$
measurement was first proposed by Xu {\it et al}.\cite{Xu} and
demonstrated by Lu {\it et al}.\cite{Lu} The film thickness of
500 nm is larger than $\lambda\approx$ 200 nm, which makes
corrections to $\lambda$ due to the sample thickness
insignificant. Conventional four-lead resistivity measurements
used for determining $H_{c2}(T,\theta)$, where $\theta$ is an
angle between the applied magnetic {\bf H} and the
crystallographic $c$ axis, were performed with a rotatable probe
in a commercial Quantum Design physical property measurement
system (PPMS), in magnetic fields between 0 T and 9 T. The
superconducting critical temperature $T_{c}$ = 38.3 K (zero
resistance) and the transition width $\Delta T_{c}$ = 0.5 K were
determined from the transport measurements. Zero-field-cooling
measurements at the MPMS with $H\approx$1 Oe show $T_c$=38.0 K.
The small value of residual resistivity ratio (RRR$\approx$4)
indicates the presence of impurities, consistent with the dirty
limit.

\section {Results and Discussion}

\subsection {MFM measurements in the MgB$_{2}$ film}

\begin{figure}
\includegraphics [trim=0 0 0 9cm,clip=true,angle=0,width=9.0cm] {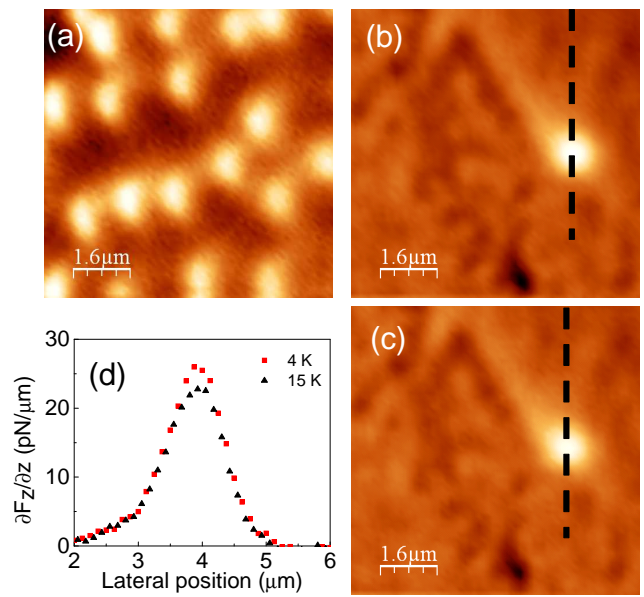}
\caption{\label{f:vortex} (Color online) (a) A typical vortex
image with a tip-lift height of 300 nm in the MgB$_2$ thin film.
(b) and (c) Single vortex images with a tip-lift height of 300 nm,
acquired at $T$ = 4 K and $T$ = 15 K, respectively. (d) The
single vortex profile along the dotted lines in (b) and (c).
Higher peak value corresponds to a smaller $\lambda$ value.}
\end{figure}

\begin{figure}
\includegraphics [trim=1cm 2cm 0 0cm,clip=true,angle=0,width=8.0cm] {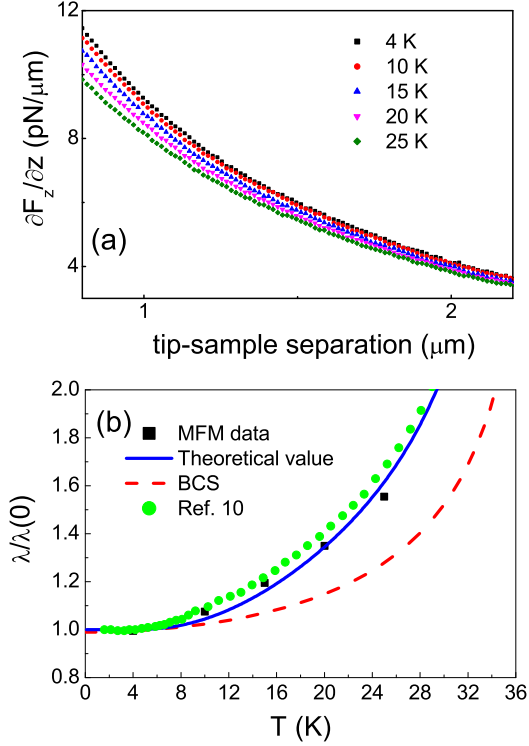}
\caption{\label{f:Meissner} (Color online) (a) Temperature
dependence of the Meissner response in MgB$_2$. (b) $\lambda(T)$
marked by the black squares are inferred from the data shown in
(a). The blue solid curve shows the calculated $\lambda(T)$ from
the gap equations for two-band superconductors. The red dashed
curve represents the conventional BCS model. The green circles
are taken from tunnel diode resonator measurements (Ref. 10).}
\end{figure}

Figure~\ref{f:vortex}(a) presents a typical vortex image in the
MgB$_2$ thin film. The well-formed vortices in the 6 $\mu$m
$\times$ 6 $\mu$m field of view were observed, which suggests the
homogeneity of the sample on a micron scale. However, the
irregular shape of individual vortices suggests the presence of
inhomogeneity in the superfluid density on a sub-micron scale,
which may be related to impurities. Figures~\ref{f:vortex}(b) and
(c) show MFM images of isolated vortices in MgB$_{2}$ at 4 K and
15 K, respectively. The features besides a single vortex
represent a sub-micron scale inhomogeneity, indicating small
variations of superfluid density. Figure~\ref{f:vortex}(d)
depicts a line profile taken along the dotted line in
Figs.~\ref{f:vortex}(b) and (c) for each of the vortices. The
maximum force gradient [$max(\partial f/\partial z)$] at the
center of the vortex qualitatively indicates that the magnitude
of $\lambda$ at 15 K is larger than that at 4 K.\cite{Auslaender
2009,Straver 2008,Shapoval 2011} In order to determine the
absolute value of $\lambda$, we performed Meissner experiments.
The force between the tip magnetic moment (a distance $d$ above
the sample) and the shielding currents induced by the tip field
is equal to the force between the real tip and the image tip,
with the mirror plane at a distance $\lambda$ below the sample's
surface.\cite{Luan 2010} This force therefore is a function of
$d+\lambda$ when $d\gg \lambda$. Direct comparison of the
Meissner curves taken at 4 K for MgB$_2$ and a reference sample
(Nb) with a known $\lambda$ gives $\lambda$(4~K) = $200 \pm 30$
nm for MgB$_2$.\cite{lambda} Comparing Meissner curves for
MgB$_2$ at 4 K and at a given temperature $T$ yields
$\delta\lambda(T)$. We obtain the absolute value of the
temperature dependent $\lambda(T)$ by adding $\delta\lambda(T)$
to $\lambda$(4 K). Figure~\ref{f:Meissner}(a) shows the Meissner
force response as a function of the tip-sample distance at
several temperatures. The systematic evolution of the Meissner
response with respect to temperature reflects the change of
$\lambda$ with temperature. Figure ~\ref{f:Meissner}(b) shows the
normalized $\lambda(T)$ (black squares) obtained for MgB$_{2}$
using the procedure outlined above, deviating significantly from
the BCS theory curve (the red dashed line), which is consistent
with the previous studies shown as green solid
circles.\cite{Fletcher} This discrepancy indicates a profound
effect of two-band superconductivity in MgB$_{2}$.\cite{Fletcher}
The large $\lambda$ in MgB$_{2}$ may be due to inclusion of
impurities, such as C, N, and Al, which significantly affects the
electron mean-free path in each band of MgB$_{2}$.

\begin{figure}
\includegraphics [trim=1cm 2cm 0 0cm,clip=true,angle=0,width=7.5cm] {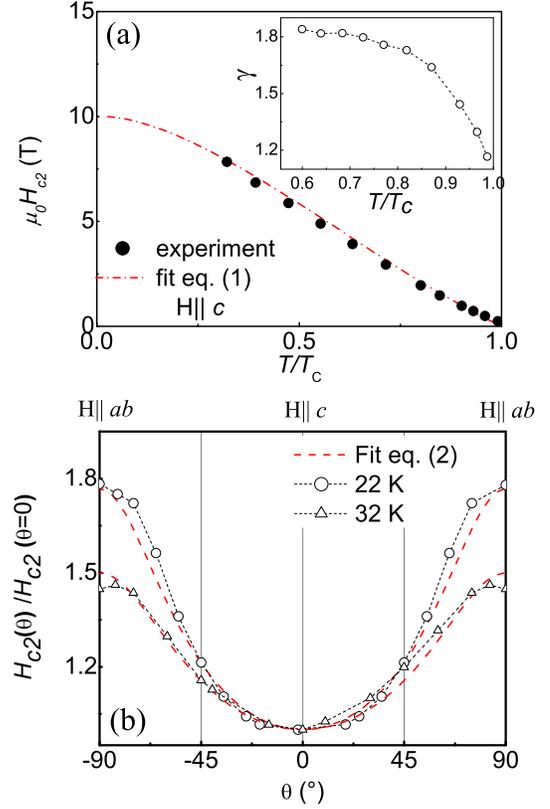}
\caption{\label{f:hc2} (Color online) (a) Numerical fit to
$H_{c2}(T)$ obtained from transport data. The inset shows the
temperature dependence of the anisotropy of $H_{c2}$. (b)
Numerical fit to $H_{c2}(\theta)$ at 22 K and 32 K with the same
parameters used to fit $H_{c2}(T)$. From the fit, the diffusivity
values of $D^{ab}_{1}=$2.36 cm$^{2}$/s and $D^{ab}_{2}=$19.7
cm$^{2}$/s were obtained; the coupling parameters obtained from
the fit are $L_{\sigma\sigma}\approx 0.810$, $L_{\pi\pi}\approx
0.285$, $L_{\sigma\pi}\approx 0.25$, and $L_{\pi\sigma}\approx
0.18$, respectively, close to the values obtained from ab-initio
calculations (Ref. 34). The uncertainty of the fit parameters is
no more than 5$\%$, which is smaller than our experimental errors
of 10$\%$.}
\end{figure}

\subsection {H$_{c2}$ measurements in the MgB$_{2}$ film}

In order to investigate the nature of disorder, we performed
temperature dependent $H_{c2}$ measurements. Figure~\ref{f:hc2}(a)
shows $H_{c2}(T)$ with the field parallel to the $c$ axis
$H^{\parallel c}_{c2}(T)$ (black circles). The value of
$H_{c2}(0)$ is considerably higher than that found in clean single
crystals ($H^{\parallel c}_{c2}(0)\approx$ 3-5 T )\cite{Perkins},
which indicates that the film is in the dirty limit. The Gurevich
model for two-band superconductors\cite{Gurevich2003} considers
inter- and intra-band scattering by non-magnetic impurities in
the dirty limit. The high $T_{c}$ in our film (which shows
essentially no suppression compared to the clean crystals) is
consistent with a small inter-band scattering, so we can use the
equations obtained for $H_{c2}(T)$ neglecting the inter-band
scattering:

\begin{equation}
\begin{split}
a_{2}[\text{ln}(t)+U(\eta h)]+a_{1}[\text{ln}(t)+U(h)]\\
+a_{0}[\text{ln}(t)+U(h)][\text{ln}(t)+U(\eta h)]=0, \label{eq:H-T} \\
\end{split}
\end{equation}
where $U(x)=\psi(1/2+x)-\psi(1/2)$, $\psi (x)$ is the di-gamma
function, $a_{1}=1+L_{-}/L_{0}$, $a_{2}=1-L_{-}/L_{0}$,
$a_{0}=2w/L_{0}$, $L_{0}=\sqrt{(L^{2}_{-}+4L_{12}L_{21})}$,
$L_{\pm}=L_{11}\pm L_{22}$, $w=L_{11}L_{22}-L_{12}L_{21}$,
$t=T/T_{c}$, $\eta=D_{2}/D_{1}$, and $h=H_{c2}D_{1}/2\Phi_{0}T$.
$\Phi_{0}$ is a single magnetic flux quantum, and $D_{1}$ and
$D_{2}$ are the intraband diffusivities. The angular-dependent
diffusivities $D_{1}(\theta)$ and $D_{2}(\theta)$ for both bands
are calculated using the following equation:

\begin{equation}
D_{m}(\theta)=\sqrt{D^{(a)2}_{m}\cos^{2}\theta+
D^{(a)}_{m}D^{(c)}_{m}\sin^{2}\theta}. \label{eq:H-Theta}
\end{equation}

The many body effects such as mass renormalization and impurity
scattering are encoded in the diffusion constants in this model.
From equations (\ref{eq:H-T}) and (\ref{eq:H-Theta}), we can
obtain $H_{c2}(T,\theta)$.

The diffusivity $D^{c}_{1}$ along the $c$ axis is smaller than the in-plane diffusivity $D^{ab}_{1}$ in
MgB$_{2}$ due to the nearly 2D nature of the $\sigma$ band. On the other hand, the values of $D^{c}_{2}$
and $D^{ab}_{2}$ do not differ substantially because of the isotropic 3D nature of the $\pi$ band. The
resulting relations among diffusivities are $D^{c}_{1} \ll D^{ab}_{1}$ and $D^{c}_{2} \approx
D^{ab}_{2}$, which leads to the anomalous behavior of the anisotropy of $H_{c2}(T)$.  The in-plane
diffusivity ratio $D^{ab}_{1}/D^{ab}_{2}$ is an important parameter in the equation (\ref{eq:H-T}).

We performed a numerical fit to three sets of transport data such
as $H_{c2}(T)$ at $\theta=$0$^{\circ}$, $H_{c2}(\theta)$ at $T =
22$ K, and $T = 32$ K using the equations (1) and (2), shown in
Fig.~\ref{f:hc2}. The relation between the best fit intraband
diffusivities in the $\sigma$ and $\pi$ bands is
$D^{a}_{2}=8.5\times D^{a}_{1}$. This large $\eta = 8.5$ is
consistent with the absence of a sharp upward curvature in
$H^{\parallel c}_{c2}(T)$  at low $T$ (see Fig. 1 in Ref. [2]),
frequently observed in C-doped MgB$_{2}$ with extremely high
$H_{c2}$. The inset of Fig.~\ref{f:hc2}(a) shows the anisotropy
$\gamma_{H_{c2}}(T)$ as a function of $T$. Again, this behavior
is qualitatively consistent with that expected for $\eta\gg 1$,
see Fig. 3(c) in Ref. [2]. The superconducting critical field,
$H^{\parallel c}_{c2}$(0), for field applied parallel to the $c$
axis, obtained from the fit, equals 10 T. This indicates the
presence of strong multiple intraband scattering channels. The
value of in-plane intraband diffusivity ratio $\eta=8.5$ provides
information about the type of the intraband scatterers. The
larger value of $\eta$, (smaller value of $D^{a}_{1}$) indicates
the weakening of the 2D $\sigma$ band by certain types of
impurities, such as C and N. These impurities affect the 2D
landscape by replacing $p_{xy}$ orbitals of boron, and making the
system more isotropic. The large value of D$^{a}_{2}$ compared to
D$^{a}_{1}$ is also in good agreement with results from the
$\alpha$ model,\cite{Fletcher} and is the result of a large
contribution of the $\pi$ band to the total density of states.

\subsection {$\lambda$ and H$_{c}$ from the two-band model}

We calculated $\lambda$ using the parameters obtained from the
$H_{c2}(T,\theta)$ fit and the band calculations. The London
equation for a two-gap superconductor is given by
$\nabla\times(\lambda^2_{L}\nabla\times\mathbf{H})+\mathbf{H}=0$,
where the London penetration depth is $\lambda_{L}^{-2}(T)=\pi
e^{2}\mu_{0}(N_{1}D^{ab}_{1}\Delta_{1}\tanh\frac{\Delta_1}{2T}+N_{2}D^{ab}_{2}\Delta_{2}\tanh\frac{\Delta_2}{2T})$:
Indices of 1 and 2 represent the $\sigma$ band and the $\pi$
band, respectively. N$_1$ and N$_2$ are the electron densities of
states. $\Delta_{1}$ and $\Delta_{2}$ are the gap magnitudes.
$D^{ab}_{1}$ and $D^{ab}_{2}$ are the intraband
diffusivities.\cite{Gurevich2003} Using $D^{ab}_{1}=$2.36
cm$^{2}$/s, $D^{ab}_{2}=19.7$ cm$^{2}$/s, obtained from the fit of
$H_{c2}(T,\theta)$, $\Delta_{1}(0)=84$ K, $\Delta_{2}(0)=33$ K,
obtained from the gap Eqs. (3), $N_1=0.3 \rm{\ states/a^3 eV}$,
and $N_2=0.41 \rm{\ states/a^3 eV}$ with a unit cell volume of
a$^3$= 87.2 \AA$^3$\cite{xxx1} (obtained from the band
calculations),\cite{Golubov} we obtain $\lambda_{L}(0)=$170$\pm$10
nm, consistent with the measured value of
$\lambda_{ab}(0)=$200$\pm$30 nm. The calculated $\lambda_{L}(T)$
is shown as the blue curve in Fig.~\ref{f:Meissner}(b),
consistent with the MFM experiment. This indicates that the two
independent measurements of $\lambda(T)$ (MFM) and $H_{c2}(T)$
(transport) in MgB$_{2}$ are complementary for investigating
superconducting properties.

\begin{figure}
\includegraphics [trim=1cm 0 0 12cm,clip=true,angle=0,width=9.0cm] {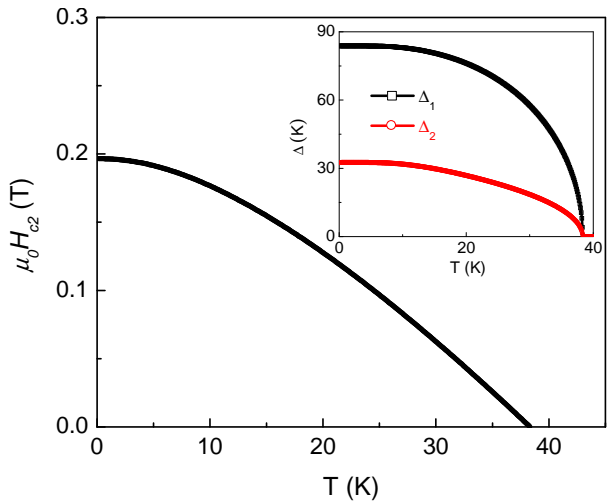}
\caption{\label{f:Hc} (Color online) The calculated thermodynamic
critical field $H_{c}$ from the gap equations for two-band
superconductors. The inset shows the calculated gap values from
the two band model.}
\end{figure}

The thermodynamic critical field ($H_{c}$) in MgB$_{2}$ is
important for technological applications.\cite{Collings} We
evaluate $H_{c}$ using the band coupling parameters, obtained from
$H_{c2}(T,\theta)$, and the electron density of states obtained
from the band calculations.

The gap equations for two-band superconductors\cite{Lin} are
\begin{equation}
\hat{g}\left(
\begin{array}{c}
  \Delta _1 \\
  \Delta _2
\end{array}
\right)-\left(
\begin{array}{c}
 N_1(0)\Delta _1Y\left(\Delta _1\right) \\
 N_2(0)\Delta _2Y\left(\Delta _2\right)
\end{array}
\right)=0,
\end{equation}
with $Y\left(\Delta _j\right)=\int _0^{\omega _c}d\xi
\frac{1}{\sqrt{\left(\xi ^2+\left|\Delta
_j\right|{}^2\right)}}\tanh \left[\frac{\sqrt{\xi ^2+\left|\Delta
_j\right|{}^2}}{2k_BT}\right]$, where $\hat{g}$ is the
superconducting coupling matrix with $g_{11}=N_1 L_{22}/w,
g_{12}=g_{21}=N_1 L_{12}/w=N_2 L_{21}/w$, and $g_{22}=N_2
L_{11}/w$. $\omega _c$ is some unknown cutoff frequency obtained
from Eqs. (3) using the $T_c$ obtained from the transport data.
Using the parameters obtained from the fit of $H_{c2}(T,\theta)$,
we have the superconducting coupling matrix, $\hat{g}=\left(
\begin{array}{cc}
 0.46 & -0.40 \\
 -0.40 & 1.78
\end{array}
\right)/\left(a^3 \text{eV}\right)$. The free energy is
calculated\cite{Lin} as $\mathcal{F}=\sum
_{\text{ij}}\left(\Delta _ig_{\text{ij}}\Delta
_j^*\right)-\frac{4}{\beta }\sum _iN_i\int _0^{\omega _c}d\xi \ln
\left(\frac{\cosh \left(\frac{1}{2}\beta \sqrt{ \left| \Delta
_i\right|{}^2+\xi ^2}\right)}{\cosh \left(\frac{1}{2}\beta \xi
\right)}\right)$. Then \(H_c\) is given by
$H_c^2/8\pi=-\mathcal{F}$. We calculate $\Delta _1(T)$ and
$\Delta _2(T)$ as shown in Fig.~\ref{f:Hc} (inset). The
calculated gap values at zero temperature are $\Delta_{1}(0)=84$
K and $\Delta_{1}(0)=33$ K, which are slightly larger than
reported values.\cite{Fletcher} The thermodynamic critical field
at zero temperature, calculated from the two-band model, is
approximately 2000 Oe. This value is smaller than those
previously obtained in polycrystalline MgB$_{2}$ by specific heat
measurements\cite{Wang} and the values reported in clean single
crystals.\cite{Bouquet,Zehetmayer}

As discussed earlier, the superconducting properties in multiband
superconductors are affected by the interactions among the
bands.\cite{Bouquet} We obtain $\xi_{ab}(0) = 5.7$ nm using
$H_{c2}(0)=\Phi_{0}/2\pi\xi^{2}(0)$ and our experimental value
$H^{\parallel c}_{c2}(0)=$ 10 T. We can then use the
Ginzburg-Landau theory to estimate the thermodynamic critical
field in the film, $H_{c}=\Phi_{0}/2\sqrt{2}\pi\lambda(0)\xi(0)=
2100 \pm 300$ Oe. This value is close to the calculated value of
$H_{c}=2000$ Oe from the two band model. This suggests that the
strong intraband scattering in the 3D $\pi$ band makes the system
more isotropic, and thus the system shows single band
characteristics.

\section {Conclusion}

In conclusion, we have measured $\lambda_{ab}(T)$ and
$H_{c2}(T,\theta)$ in a MgB$_2$ film. Our analysis of
$H_{c2}(T,\theta)$ shows that the large value of the in-plane
intra-band diffusivity in the 3D $\pi$ band is due to the presence
of non-magnetic impurities such as C and N, indicating the system
is more isotropic, which is partly responsible for a large
$\lambda$. We calculated $\lambda$ and $H_{c}$ employing the gap
equations for the two-band superconductors using the parameters
obtained from $H_{c2}(T,\theta)$ and derived from band
calculations. The calculated $\lambda_{L}(0)=$170$\pm$10 nm is
close to the measured $\lambda(0) = 200\pm 30$ nm in MgB$_2$ film,
indicating that two independent measurements, such as MFM and
transport, are complementary, and provides a venue for thoroughly
investigating superconducting properties. The determination of
$H_{c}(T)$ in clean MgB$_{2}$ and in multi-band superconductors
in general is a fascinating problem with both fundamental and
technological relevance.

We acknowledge valuable discussions and communication of data
with A. Gurevich. This work was supported by the US Department of
Energy, Basic Energy Sciences, Division of Materials Sciences and
Engineering, at Los Alamos. Work at Brookhaven was supported by
the US Department of Energy under Contract No. DE-AC02-98CH10886.
N.H. is member of CONICET (Argentina).

\end{document}